# A New Era in Citation and Bibliometric Analyses: Web of Science, Scopus, and Google Scholar

## Lokman I. Meho[*] and Kiduk Yang

School of Library and Information Science, Indiana University - 1320 East 10$^{th}$ St., LI 011; Bloomington, IN 47405; Tel: 812-855-2018; meho@indiana.edu; kiyang@indiana.edu

**Abstract:** Academic institutions, federal agencies, publishers, editors, authors, and librarians increasingly rely on citation analysis for making hiring, promotion, tenure, funding, and/or reviewer and journal evaluation and selection decisions. The Institute for Scientific Information's (ISI) citation databases have been used for decades as a starting point and often as the only tools for locating citations and/or conducting citation analyses. ISI databases (or *Web of Science*), however, may no longer be adequate as the only or even the main sources of citations because new databases and tools that allow citation searching are now available. Whether these new databases and tools complement or represent alternatives to *Web of Science* (*WoS*) is important to explore. Using a group of 15 library and information science faculty members as a case study, this paper examines the effects of using *Scopus* and *Google Scholar* (*GS*) on the citation counts and rankings of scholars as measured by *WoS*. The paper discusses the strengths and weaknesses of *WoS*, *Scopus*, and *GS*, their overlap and uniqueness, quality and language of the citations, and the implications of the findings for citation analysis. The project involved citation searching for approximately 1,100 scholarly works published by the study group and over 200 works by a test group (an additional 10 faculty members). Overall, more than 10,000 citing and purportedly citing documents were examined. *WoS* data took about 100 hours of collecting and processing time, *Scopus* consumed 200 hours, and *GS* a grueling 3,000 hours.

## INTRODUCTION

Because it is sometimes hard for funding agencies, administrators, tenure and promotion committees, and others to rely only on peer reviews and publication output for measuring the impact or quality of an author's work, they additionally use citation analysis to perform such evaluations. In general, the use of citations for evaluating research is based on the assumption that citation counts are an objective measure that credits and recognizes the value, impact, quality, or significance of an author's work (Borgman & Furner, 2002; Cronin, 1984; Holden, Rosenberg, & Barker, 2005; Moed, 2005; van Raan, 1996, 2005; Wallin, 2005).

---

[*]To whom all correspondence should be addressed.

Many scholars have argued for and against the use of citations for assessing research quality. Proponents have reported the validity and reliability of citation counts in research assessments as well as the positive correlation between these counts and peer reviews/assessments of publication venues (Aksnes & Taxt, 2004; Glänzel, 1996; Holmes & Oppenheim, 2001; Kostoff, 1996; Martin, 1996; Narin, 1976; Oppenheim, 1995; So, 1998; van Raan, 2000; Warner, 2000). Critics, on the other hand, claim that citation counting has serious problems or limitations that affect its validity (MacRoberts & MacRoberts, 1996; Seglen, 1998). Important limitations reported in the literature focus on, among other things, the problems associated with the data sources used, especially the Institute for Scientific Information (ISI, currently Thomson Scientific) citation databases: *Arts & Humanities Citation Index*, *Science Citation Index*, and *Social Sciences Citation Index*—the standard and most widely used tools for generating citation data for research and other assessment purposes. These tools are now currently part of what is known as *Web of Science* (*WoS*), the portal used to search the three ISI citation databases. In this paper, we use ISI citation databases and *WoS* interchangeably.

Critics of ISI citation databases note that they: (1) cover mainly North American, Western European, and English-language titles; (2) are limited to citations from 8,700 journals;[1] (3) do not count citations from books and most conference proceedings; (4) provide different coverage between research fields; and (5) have citing errors, such as homonyms, synonyms, and inconsistency in the use of initials and in the spelling of non-English names (many of these errors, however, come from the primary documents themselves rather than being the result of faulty ISI indexing).

---

[1]*Ulrich's Periodicals Directory* lists approximately 22,500 active academic/scholarly, refereed journals. Of these, approximately 7,500 are published in the United States, 4,150 in the United Kingdom, 1,600 in the Netherlands, 1,370 in Germany, 660 in Australia, 540 in Japan, and 500 in Canada, 450 in China, 440 in India, and 420 in France.

Studies that have addressed problems of, and/or suggested alternative or complementary sources to, ISI citation databases are very few and can be divided into two groups:

- Studies that examined the effect of certain limitations in ISI database by comparing its coverage with that of other citation databases or tools; and

- Studies that suggested or explored different or additional sources and methods for identifying citations.

**Studies that Examined the Effect of Certain Limitations in ISI Databases**

In a study aimed at analyzing the effect of the omission of certain journals in ISI databases on citation-based appraisals of communication literature, Funkhouser (1996) examined references in 27 communication journals (13 covered by ISI and 14 not covered) for the year of 1990. He found that 26% of author citations were from non-ISI journals and that 27 of the 50 most highly cited authors received at least 25% of their citations from non-ISI journals. Funkhouser, however, did not verify whether the omission of those 14 journals had any impact on the relative citation ranking of scholars if one relied only on ISI data.

Cronin, Snyder, and Atkins (1997) analyzed thousands of references from monographs and leading academic journals in sociology to identify the effects of ISI databases' non-coverage of citations in monographic literature. They found that the relative rankings of authors who were highly cited in the monographic literature did not change in the journal literature of the same period. The overlap of citations in monographs and journals, however, was small, suggesting that there may be two distinct populations of highly cited authors.

Whitley (2002) compared the duplication and uniqueness of citing documents in *Chemical Abstracts* and *Science Citation Index* for the works of 30 chemistry researchers for the years 1999–2001. She found that 23% of all the citing documents were unique to *Chemical Abstracts*, 17% were unique to the *Science Citation Index*, and the remaining 60% were duplicated in the two databases. Whitley concluded that relying on either index alone would lead to faulty results when trying to obtain citation totals for individual authors.

Goodrum et al. (2001) and Zhao and Logan (2002) compared citations from *CiteSeer/ResearchIndex*, a Web-based citation indexing system, with those from ISI's *Science Citation Index* (*SCI*) in the field of computer science. Both studies found a 44.0% overlap among the top-25 cited authors and concluded that *CiteSeer/ResearchIndex* and *SCI* were complementary in their coverage of the field. More recently, Pauly and Stergiou (2005) compared citation counts between *WoS* and *GS* for papers in mathematics, chemistry, physics, computing sciences, molecular biology, ecology, fisheries, oceanography, geosciences, economics, and psychology. Each discipline was represented by three authors, and each author was represented by three articles (i.e., 99 articles in total). The authors also examined citations to an additional 15 articles for a total of 114. Without assessing the accuracy or relevance and quality of the citing articles, the authors reported such good correlation in citation counts between the two sources that they suggested *GS* can substitute for *WoS*.

Bauer and Bakkalbasi (2005) compared citation counts provided by *WoS, Scopus,* and *Google Scholar* (*GS*) for articles from the *Journal of the American Society for Information Science and Technology* published in 1985 and in 2000. They found that *WoS* provided the highest citation counts for the 1985 articles and *GS* provided statistically significant higher citation counts than either *WoS* or *Scopus* for the 2000 articles. They did not find significant differences between *WoS* and *Scopus* for either year. The authors, however, stated that more rigorous studies were required before these findings could be considered definitive, especially because the scholarly value of some of the unique material found in *GS* remained an open question.

Jacsó (2005a) also conducted several tests comparing *GS*, *Scopus,* and *WoS*, searching for documents citing: (1) Eugene Garfield; (2) an article by Garfield published in 1955 in *Science*; (3) the journal *Current Science*; and (4) the 30 most cited articles from *Current Science*. He found that coverage of *Current Science* by *GS* is "abysmal" and that there is considerable overlap between *WoS* and *Scopus*. He also found many unique documents in each source, claiming that the majority of the unique items were relevant and substantial. For "lack of space," Jacsó's analysis was limited to reporting citation counts and retrieval performance by time period; he did not provide an in-depth analysis and examination of, for example, the type, refereed status, and source of the citations.

Most recently, Bakkalbasi, Bauer, Glover, and Wang (2006) compared citation counts for articles from two disciplines (oncology and condensed matter physics) and two years (1993 and 2003) to test the hypothesis that the different scholarly publication coverage provided by *WoS*, *Scopus*, and *GS* would lead to different citation counts from each. They found that for oncology in 1993, *WoS* returned the highest average number of citations; 45.3, *Scopus* returned the highest average number of citations (8.9) for oncology in 2003; and *WoS* returned the highest number of citations for condensed matter physics in 1993 and 2003 (22.5 and 3.9 respectively). Their data showed a significant difference in the mean citation rates between all pairs of resources except between *Scopus* and *GS* for condensed matter physics in 2003. For articles published in 2003 *WoS* returned the largest amount of unique citing material for condensed matter physics and *GS* returned the most for oncology. Bakkalbasi, Bauer, Glover, and Wang concluded that all three tools returned some unique material and that the question of which tool provided the most complete set of citing literature might depend on the subject and publication year of a given article.

**Studies that Suggested Sources and Methods beyond ISI or Citation Databases**

In a 1995 paper Reed recommended that faculty seeking tenure or promotion: (1) review the citations in selected key journals in their specialty that were not covered in ISI databases; (2) scan the citations and bibliographies in textbooks and monographs pertinent to their research areas; (3) record citations discovered through research, teaching activities, and professional reading throughout their careers; and (4) maintain a continuously updated file of citations as they are discovered. These recommendations were adopted by Nisonger (2004a), who additionally suggested that (5) sources be examined (e.g., books, journal articles, and doctoral dissertations identified in major bibliographies in one's specialty area), and that (6) the author's name be searched on the Web for items not indexed in ISI databases.

Unlike Reed who only compiled and recommended a list of techniques to locate citations not covered by ISI, Nisonger conducted a self-study to show how ISI coverage compared to citation data he collected using the aforementioned six techniques. His study was based on analysis of his own lifetime citation record which he compiled by (a) searching the ISI databases; (b) manually searching the literature for nearly 15 years; and (c) making use of various Web search engines. He found that (with self-citations excluded) ISI captured 28.8% of his total citations, 42.2% of print citations, 20.3% of citations from outside the United States, and 2.3% of non-English citations. Nisonger suggested that faculty should not rely solely on ISI author citation counts, especially when demonstration of international impact is important. He also suggested that rankings based on ISI data of a discipline's most-cited authors or academic departments might be significantly different if non-ISI citation data were included. This suggestion, however, was not verified by empirical data; it merely suggested that broader sourcing of citations might alter one's relative ranking vis-à-vis others.

**Research Questions and Significance**

While both Reed's recommendations and Nisonger's methods are useful techniques for locating citations, they are not practical in the case of large study samples. Citation databases remain the most viable methods for generating bibliometric data and for making accurate citation-based research assessments and large-scale comparisons between works, authors, and journals. Until very recently, *WoS* was the standard tool for conducting extensive citation searching and bibliometrics analysis, primarily because it was the only general and comprehensive citation database in existence. This, however, may no longer be the case because several databases or tools that provide citation searching capabilities have appeared in the past few years (see below). Some of these databases or tools are sufficiently comprehensive and/or multidisciplinary in nature such that they pose a direct challenge to the dominance of *WoS* and raise questions about the accuracy of using it exclusively in citation, bibliometric, and scholarly communication studies. Thus, several research questions suggest themselves:

1) What is the impact of using new, additional citation databases or tools on the counting and ranking of works, authors, journals, and academic departments?

2) How do the citations generated by these new tools compare with those found in *WoS* in terms of, for example, document source, document type, refereed status, language, and subject coverage?

3) Do these tools represent alternatives or complements to *WoS*?

4) What strengths and weaknesses do these new citation tools have relative to *WoS*?

Answering these questions is important to anyone trying to determine whether an article, author, or journal citation search should be limited to *WoS*. The answers to these questions are also important for those seeking to use appropriate tools to generate more precise citation counts, rankings, and assessments of research impact than those based exclusively on *WoS*. More complete citation counts can help support or identify more precisely discrepancies between research productivity, peer evaluation, and citation data. More complete citation counts can also help generate more accurate $h$-index scores for authors and journals (Bornmann & Daniel, 2005, 2007; Hirsch, 2005) and journal impact factors (Garfield, 1996, 2006; Nisonger, 2004b; Saha, Saint, & Christakis, 2003), as well as identify international impact (Nisonger, 2004a; de Arenas, Castanos-Lomnitz, & Arenas-Licea, 2002).

Scholars trying to locate citations to a specific publication for traditional research purposes (as opposed to citation counts, research evaluation, and so on) will find answers to the aforementioned questions very useful, too, especially in cases where bibliographic searches fail to identify relevant research materials. Serials librarians who use citation counts and analyses to make journal subscription and cancellation decisions will benefit from the findings of this study as well since it has the potential to show whether there is a need to rely on multiple sources of citation data. Vendors and producers of full-text databases, such as CSA, EBSCO, Elsevier, OCLC, Ovid, ProQuest, Sage, Springer, Taylor & Francis, and Wilson will also benefit from answers to these questions by applying the findings of this study to develop and illustrate additional features and uses of their databases.

**Competitors to *Web of Science***

Today, there are more than 100 databases or tools that allow citation searching. These databases or tools can be classified into three basic categories. The first allows the user to search in the full text field to determine whether certain papers, books, authors or journals have been cited in a document. Examples of these databases or tools include *ACM Digital Library*, *arXiv.org*, *Emerald Full Text*, *ERIC*, *Google Book Search*, *IEEE Computer Society Digital Library* and *IEEE Xplore*, *Library Literature and Information Science Full Text*, *NetLibrary*, and Elsevier's *Scirus*. Also belonging to this category are databases or tools that automatically extract and parse bibliographic information and cited references from electronic full text documents retrieved from personal homepages and digital archives and repositories. Examples of these include *CiteSeer* (computer science), *Google Scholar* (general), *RePEc* (economics), and *SMEALSearch* (business).

The second category of databases or tools allows the user to search in the cited references field to identify relevant citations. Examples of these include: *Library, Information Science & Technology Abstracts*, *PsycINFO*, *PubMed Central*, and Elsevier's *ScienceDirect*. The last category is databases that serve exactly like *Web of Science*. The main and perhaps only good example of this category is *Scopus*.

Details about the citation searching features and strengths and weaknesses of these and many other databases that allow citation searching can be found in Roth (2005), Ballard and Henry (2006), and the many review and scholarly articles by Peter Jacso (see: http://www2.hawaii.edu/~jacso/).

Although there are many databases and services that could be used to answer the above-mentioned research questions, this study focuses on:

1. Analyzing the effects of using *Scopus* and *GS* on the citation counts and rankings of individual scholars as measured by *WoS*, using a group of library and information science (LIS) faculty members as a case study. LIS makes an ideal case study due to the interdisciplinary and multidisciplinary nature of its research areas and its use of, and reliance on, various types of literature for scholarly communication (e.g., journal articles, conference papers, and books).

2. Examining the similarities and differences between *WoS*, *Scopus*, and *GS* in terms of coverage period, sources of citations, document type, refereed status, language, and subject coverage, and identifying strengths and weaknesses of the three tools; and

3. Discussing the implications of the findings on citation analysis and bibliometric studies.

*Scopus* and *GS* were chosen because of their similarity to *WoS* in that they were created specifically for citation searching and bibliometric analysis, in addition to being useful for bibliographic searches. *Scopus* and *GS* were also chosen because they represent the only real or potential competitors to *WoS* in citation analysis and bibliometrics research areas. More information about these three sources is provided below.

## METHODS

**Citation Databases or Tools**

*WoS*, which comprises the three ISI citation databases (*Arts & Humanities Citation Index*, *Science Citation Index*, and *Social Sciences Citation Index*), has been the standard tool for a significant portion of citation studies worldwide. A simple keyword search in *WoS* and other databases (e.g., *Library and Information Science Abstracts*, *Pascal*, *Medline*, *EMBASE*, *Biosis Previews*, and *INSPEC*) indicates that ISI databases have been used, or referred to, in several thousand journal articles, conference papers, and chapters in books published in the last three decades. *WoS*'s website provides substantial factual information about the database, including the number of records and the list of titles indexed. It also offers powerful features for browsing, searching, sorting, and saving functions, as well as exporting to citation management software (e.g., *EndNote* and *RefWorks*). Coverage in *WoS* goes back to 1900 for *Science Citation Index*, 1956 for *Social Sciences Citation Index*, and 1975 for *Arts & Humanities Citation Index*. As of October 2006, there were over 36 million records in the database (the version the authors had access to) from approximately 8,700 scholarly titles (Thomson Corporation, 2006a), including several hundred conference proceedings and over 190 open access journals (Harnad & Brody, 2004).[2] Over 100 subjects are covered in *WoS*, including all the major arts, humanities, sciences, and social sciences subdisciplines (e.g., architecture, biology, business, chemistry, health sciences, history, medicine, political science, philosophy, physics, religion, and sociology). For more details on *WoS*, see Goodman and Deis (2005) and Jacsó (2005a).

---

[2]The figure for conference proceedings was generated by analyzing the source titles of over 125,000 records that were published in the *Lecture Notes* series (e.g., *Lecture Notes in Artificial Intelligence, Lecture Notes in Computer Science*, and *Lecture Notes in Mathematics*). Also analyzed were the indexed titles that included the word conference, proceedings, symposium, workshop, or meeting in their names.

Similar to ISI, Elsevier, the producer of *Scopus*, provides substantial factual information about the database, including the number of records and the list of titles indexed. It also offers powerful features for browsing, searching, sorting, and saving functions, as well as exporting to citation management software. Coverage in *Scopus* goes back to 1966 for bibliographic records and abstracts and 1996 for citations. As of October 2006, there were over 28 million records in the database from over 15,000 "peer-reviewed" titles, including coverage of 500 Open Access journals, 700 conference proceedings, 600 trade publications, and 125 book series (Elsevier, 2006). Subject areas covered in *Scopus* include: Chemistry, Physics, Mathematics, and Engineering (4,500 titles), Life and Health Sciences (5,900 titles, including 100% Medline coverage), Arts and Humanities, Social Sciences, Psychology, and Economics (2,700 titles), Biological, Agricultural, and Environmental Sciences (2,500 titles), and General Sciences (50 titles). For more details on *Scopus*, see Goodman and Deis (2005) and Jacsó (2005a).

In contrast to ISI and Elsevier, Google does not offer a publisher list, title list, document type identification, or any information about the time-span or the refereed status of records in *GS*. This study, however, found that *GS* covers print and electronic journals, conference proceedings, books, theses, dissertations, preprints, abstracts, and technical reports available from major academic publishers, distributors, aggregators, professional societies, government agencies, and preprint/reprint repositories at universities, as well as those available across the web. Examples of these sources include: Annual Reviews, arXiv.org, Association for Computing Machinery (ACM), Blackwell, Cambridge Scientific Abstracts (CSA), Emerald, HighWire Press, Ingenta, Institute of Electrical and Electronics Engineers (IEEE), PubMed, Sage, Springer, Taylor & Francis, University of Chicago Press, and Wiley, among others (Bauer & Bakkalbasi, 2005; Gardner & Eng, 2005; Jacsó, 2005b; Noruzi, 2005; Wleklinski, 2005). Although *GS* does not cover material from all major publishers (e.g., American Chemical Society and Elsevier), it identifies citations to articles from these publishers when documents from other sources cite these articles. *GS* does not indicate how many documents it searches.

Table 1 provides detailed information about the breadth and depth of coverage, subject coverage, citation browsing and searching options, analytical tools, and downloading and exporting options of all three sources.

**Units of Analysis**

In order to analyze the effect of using additional sources to *WoS* on the citation counts and rankings of LIS faculty members and to be able to generalize the findings to the field, this study explored the difference *Scopus* and *GS* make to results from *WoS* for all 15 faculty members of the School of Library and Information Science at Indiana University-Bloomington (SLIS).[3] These faculty members cover most of the mainstream LIS research areas as identified by the Association of Library and Information Science Education (ALISE, 2006); they also cover research areas beyond those listed by ALISE (e.g., computer-mediated communication and computational linguistics). Moreover, SLIS faculty members are the most published and belong to one of the most cited American Library Association accredited LIS programs in North America (Adkins & Budd, 2006; Budd, 2000; Persson & Åström, 2005; Thomson Corporation, 2006b). From 1970 to December 2005, the 15 SLIS faculty members had published or produced over 1,093 scholarly works, including: 312 refereed journal articles, 12 refereed review articles, 305 conference papers (almost all refereed), 131 chapters (some refereed), 36 books, and 35 edited volumes, among others (see Table 2).

All data were entered into EndNote libraries and Access databases and were coded by citing source (e.g., journal name, conference, book, and so on), document type (e.g., journal article, review article, conference paper, and so on), refereed status of the citing item, year, language, and source used to identify the citation. The refereed status of the sources of citations was determined through *Ulrich's International Periodicals Directory* and the domain knowledge of the researchers and their colleagues.

---

[3]As of January 2007, SLIS had 17 full-time faculty members.

**Data Collection**

All *WoS* and *Scopus* data were manually collected and processed twice by one of the authors (LIM) in October 2005 and again (for accuracy and updating purposes) in March 2006. *GS* data were harvested in March 2006; however, identifying their relevancy and full bibliographic information took approximately 3,000 hours of work over a six months period, which included manually verifying, cleaning, formatting, standardizing, and entering the data into EndNote Libraries and Access databases.[4]

The "Cited Author" search option was used in *WoS* to identify citations to each of the 1,093 items published by the 15 faculty members constituting the study group. Citations to items in which the faculty members were not first authors, as well as citations to dissertations and other research materials written by them, were included in the study. Although publicly available, the data have been anonymized, assigning citations to faculty members by their research areas rather than by names.

Unlike *WoS*, *Scopus* does not have browsing capabilities for the cited authors or cited works fields that would allow limiting the search to relevant citations (cited works field is the index field for names of journals, books, patent numbers, and other publications). As a result, instead of browsing the cited authors or cited works fields, we used an *exact match* search approach to identify all potentially relevant citations in the database. This method uses the title of an item as a search statement (e.g., "Invoked on the Web") and tries to locate an exact match in the cited "References" field of a record. Using the titles of the 1,093 items published or produced by the study group, the method allowed us to identify the majority of the relevant citations in the database. In cases where the title was too short or ambiguous to refer to the item in question, we used additional information as keywords (the first author's last name and, if necessary, journal name and/or book or conference title) to ensure that we retrieved only relevant citations. In cases where the title was too long, we used the first few words of the title because utilizing all the words in a long title increases the possibility of missing some relevant citations due to typing or indexing errors. When in doubt, we manually examined all retrieved records to make sure they cited the items in question. Other search options in the database were used (e.g., "Author Search" and "Advanced Search"), but they not only did not identify any unique, additional citations, they were less inclusive than the exact match approach. For example, because not all of the 1,093 items published by the study group are indexed in the database, the "Author Search" approach would have been inappropriate or would have resulted in incomplete sets of relevant citations.

---

[4]At the time of data collection, *GS* did not provide the option of downloading search results into a bibliographic management software (e.g., EndNote, BibTeX, and RefWorks). While the ability to download search results into any of these programs would have reduced the amount of time spent on processing the citations, manually verifying, cleaning, formatting, and standardizing the citations would still have been necessary and would have consumed an excessive amount of time.

*GS* was searched for citations using two methods: *Author* search and *exact match* (or *exact phrase*) search. The Author search usually retrieves items published by an author and ranks the items in a rather inconsistent way. Once the items are retrieved, a click on the "Cited by . . ." link allows the searcher to display the list of citing documents. The "Cited by . . ." link is automatically generated by *GS* for each cited item.

The exact match search approach was used to ensure that citations were not missed due to errors in *GS*'s Author search algorithm. This strategy, which is the same as applied in *Scopus*, resulted in 1,301 records. Of these, 534 were unique relevant citations. In other words, if the exact match search approach was not used, in addition to the Author search approach, 534 (or 14.6%) of *GS*'s relevant citations would have been missed. The remaining 767 records retrieved through the exact match search were either previously found through the Author search approach or were not relevant. Almost all of the false drops were documents retrieved when searching for citations to short-title items.

A major disadvantage of *GS* is that its records are retrieved in a way that is very impractical for use with large sample sizes, requiring a very tedious process of manually extracting, verifying, cleaning, organizing, classifying, and saving the bibliographic information into meaningful and usable formats. Moreover, unlike *WoS* and *Scopus*, *GS* does not allow re-sorting of the retrieved sets in any way (such as by date, author name, or data source); as mentioned earlier, retrieved records in *GS* are rank-ordered in a rather inconsistent way. The result sets show short entries, displaying the title of the cited article and the name of the author(s) and, in some cases, the source. Entries that include the link "Cited by . . ." indicate the number of times the article has been cited. Clicking on this link will take users to a list of citing articles. Users will be able to view the full-text of only those items that are available for free and those that their libraries subscribe to.

Other major disadvantages of *GS* include duplicate citations (e.g., counting a citation published in two different forms, such as preprint and journal article, as two citations), inflated citation counts due to the inclusion of non-scholarly sources (e.g., course reading lists), phantom or false citations due to the frequent inability of *GS* to recognize real matches between cited and citing items claiming a match where there is not even minimal "chemistry" (Jacsó, 2006), errors in bibliographic information (e.g., wrong year of publication), as well as the lack of information about document type, document language, document length, and the refereed status of the retrieved citations. In many cases, especially when applying the Exact Phrase search method, the item for which citations are sought is retrieved and considered a citation by *GS* (in such cases, these citations were excluded from the search results). Perhaps the most important factor that makes *GS* very cumbersome to use, is the lack of full bibliographic information for citations found. Even when some bibliographic information is made available (e.g., source), it is not provided in a standard way thus requiring a considerable amount of manual authority control, especially among citations in conference proceedings. For example, the *Annual Meeting of the American Society for Information Science and Technology* is cited in at least five different ways (ASIST 2004: …, ASIST 2005: …, Proceedings of the American Society for Information Science and Technology, Annual Meeting of the …, and so on) whereas in *WoS* and *Scopus* all entries for this conference and other conference proceedings are entered in standardized fashion. The presence of all these problems in *GS* suggest that unless a system is developed that automatically and accurately parses result sets into error-free, meaningful, and usable data, *GS* will be of limited use for large-scale comparative citation and bibliometric analyses.

To make sure that citations were not overlooked because of searching or indexing errors, we looked for the bibliographic records of all citations that were missed by one or two of the three tools. For example, if a citation was found in *WoS* but not in *Scopus* or *GS*, we conducted bibliographic searches in *Scopus* and/or *GS* to see if the item were in fact indexed in them. When the bibliographic record of any of these missed citations was found in one of the three tools, we examined: (1) why it was not retrieved through the citation search methods described above; and (2) whether it should be counted as a citation. Items that were overlooked due to searching errors (16 in the case of *WoS* and 27 in *Scopus*) were counted as citations toward their respective databases; most of the searching errors were due to having missed selecting a relevant entry when browsing the cited references field in *WoS* and making typographical errors when entering a search query in *Scopus*. Items that were missed due to database/system errors were tallied but were not counted as citations. These included:

- *WoS*: 10 citations were missed due to incomplete lists of references. These citations are the equivalent of 0.5% of the database's relevant citations.

- *Scopus*: 75 citations were missed due to lack of cited references information and 26 citations due to incomplete lists of references in their respective records. In total, *Scopus* missed 101 (or the equivalent of 4.4% of its relevant citations) due to database errors.

- *GS*: missed 501 (or the equivalent of 12.0% of its relevant citations) due to system errors. Many of the errors in *GS* were a result of matching errors. For example, the search engine failed, in many cases, to identify an exact match with the search statements used because a word or more in the title of the cited item was automatically hyphenated in the citing document. Or *GS* failed to retrieve relevant citations from documents that do not include well defined sections named a Bibliography, Cited References, Cited Works, Endnotes, Footnotes, or References.

These results suggest that if citation searching of individual LIS scholars were limited to *Scopus*, a searcher would miss an average of 4.4% of the relevant citations due to database errors. In the case of *GS*, the percentage would be 12.0%; this percentage would increase to 26.6% had we not used the Exact Phrase search approach described earlier. The results also suggest that when using *GS* one must use both the Author search and Exact Phrase search methods.

It is important to note here that it took about 100 hours of work to collect, clean, standardize, and enter all the data into EndNote libraries and Access databases from *WoS*, about 200 hours in the case of *Scopus*, and, as mentioned earlier, over 3,000 hours in the case of *GS*. In other words, collecting *GS* data took 30 as much time as collecting *WoS* data and 15 as much time as that of *Scopus*—this includes the time needed to double-check the missed items in each source. It is also important to note that in studies such as ours, it is essential that the investigators have access to complete lists of publications of the group being investigated. Without this information, there would be major problems with the data collected, especially when there are authors with common names among the study group. In our case, all members of the study group either had their complete publication information available online or they provided it on request. This information was very useful in the case of more than half of the study group as we discovered multiple authors with the names B. Cronin, S. Herring, J. Mostafa, N. Hara, D. Shaw, and K. Yang. The availability of their publication lists helped avoid including non-relevant citations.

## RESULTS AND DISCUSSION

The results of this study are presented and discussed in three sections: (1) the effect of using *Scopus* on the citation counts and rankings of SLIS faculty members as measured by *WoS*; (2) the effect of using *GS* on the citation counts and rankings as measured by *WoS* and *Scopus* combined; and (3) the sources of citations found in all three tools, including their names (i.e., journal and conference proceedings), refereed status, and language. Because the three tools provide different citation coverage in terms of document type and time period, we limited most of the analysis to citations from types of documents and years common to all three tools, that is, conference papers and journal items (e.g., journal articles, review articles, editorials, book reviews, and letters to the editor) published between 1996 and 2005. Excluded from the analysis are citations found in books, dissertations, theses, reports, and so on, as well as 475 citations from *GS* that did not have complete bibliographic information. These 475 citations primarily included: bachelor's theses, presentations, grant and research proposals, doctoral qualifying examinations, submitted manuscripts, syllabi, term papers, working papers, web documents, preprints, and student portfolios.

**Effect of *Scopus* on Citation Counts and Rankings of LIS Faculty**

To show the difference that *Scopus* makes to the citation counts and rankings of LIS faculty members as measured by *WoS*, we compare the number of citations retrieved by both databases, show the increase *Scopus* makes toward the total number of citations of SLIS as a whole and also of individual faculty members, explore the effect *Scopus* has on altering the relative citation ranking of SLIS faculty members, and examine the overlap and unique coverage between the two databases. The refereed status of citations found in *WoS* and *Scopus* is not discussed because the great majority of citations from these two databases come from scholarly, peer-reviewed journals and conference proceedings.

As shown in Tables 3 and 4, *Scopus* includes 278 (or 13.7%) more citations than *WoS*, suggesting that *Scopus* provides more comprehensive coverage of the LIS literature than *WoS*.[5] Further analysis of the data shows that combining citations from *Scopus* and *WoS* increases the number of citations of SLIS as a whole by 35.1% (from 2,023 to 2,733 citations). This means that were only *WoS* used to locate citations for LIS faculty members, on average, more than one-third of relevant citations (found in the union of *WoS* and *Scopus*) would be missed; the percentage of missed citations would be 18.8% were only *Scopus* used.

Perhaps more importantly, the data show that the percentage of increase in citation counts for individual faculty members varies considerably depending on their research areas, ranging from 4.9% to 98.9%. For example, faculty members with research strengths in such areas as communities of practice, computational linguistics, computer-mediated communication, data mining, data modeling, discourse analysis, gender and information technology, human-computer interaction, information retrieval, information visualization, intelligent interfaces, knowledge discovery, and user modeling, will find their citation counts increase considerably more than those faculty members with research strengths in other areas (see Table 4). These findings not only imply that certain subject areas will benefit more than others from using both *Scopus* and *WoS* to identify relevant citations, they also suggest that to generate accurate citation counts for faculty members, and by extension schools, and to accurately compare them to one another, a researcher must use both databases. The importance of using *Scopus* in addition to *WoS* is further evidenced by the facts that:

---

[5]Table 3 also shows that *WoS* includes 391 (or 17.0%) more citations than *Scopus* (2,692 in comparison to 2,301, respectively), when citations from pre-1996 are counted.

- The relative ranking of faculty members changes in eight out of 15 cases, strikingly so in the cases of faculty members E, F, H, and I (see Table 5). Although the overall relative ranking of the faculty members does not change significantly when citations from both databases are counted (Spearman Rank Order correlation coefficient = 0.9134 at 0.01 level), the rankings do change significantly when faculty members in the middle third of the rankings are examined separately (Spearman Rank Order correlation coefficient = -0.45 at 0.01 level). In other words, *Scopus* significantly alters the relative ranking of those scholars that appear in the middle of the rankings but not for those at the top or bottom of the rankings.

- The overlap of LIS citations between the two databases is relatively low—58.2% (see Figure 1) with significant differences from one research area to another ranging from a high 82.0% to a low 41.1% (see Table 6).

- The number of unique citations found in *Scopus* is noticeably high in comparison to that of *WoS* (710 or 26.0% in comparison to 432 or 15.8%, respectively) (see Figure 1). The overlap and uniqueness between the two databases is almost identical to what Whitley (2002) found in her study that compared the duplication (60%) and uniqueness of citing documents in *Chemical Abstracts* (23%) and *Science Citation Index* (17%).

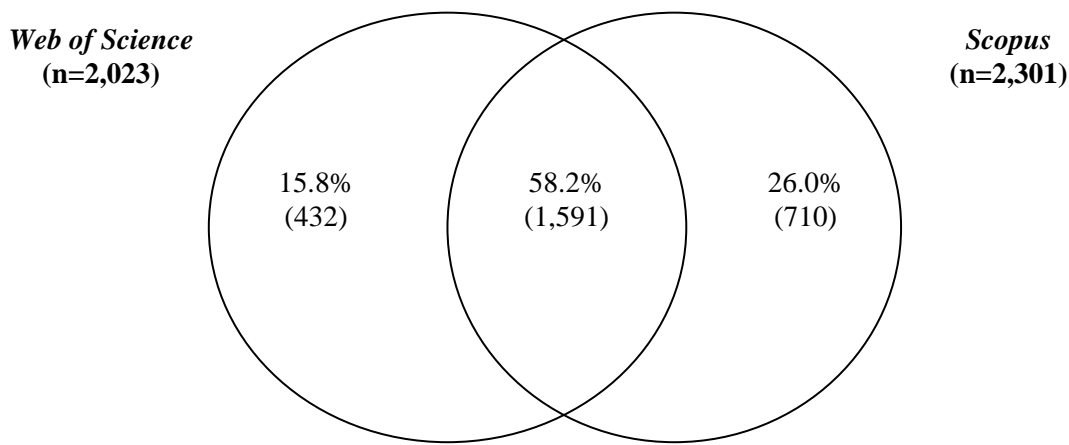

**Figure 1. Distribution of unique and overlapping citations in *WoS* and *Scopus* (N=2,733)**

Regarding the type of documents in which the citations were found, the main difference between the two databases is in the coverage of conference proceedings. *Scopus* retrieves considerably more citations from refereed conference papers than *WoS* (359 in comparison to 229, respectively) (see Table 7). What is more important is that of all 496 citations from conference papers, 53.8% are uniquely found in *Scopus* in comparison to only 27.6% in *WoS* (19.6% of citations from conference papers are found in both databases). This can have significant implications for citation analysis and the evaluation of individual scholars, especially when those evaluated include authors who use conferences as a main channel of scholarly communication. Without *Scopus*, authors who communicate extensively through conferences will be at a disadvantage when their citation counts are compared with those who publish primarily in journals due to poor coverage of conference proceedings in *WoS*. Whether the value, weight, or quality of citations found in conference papers is different from those of journal articles is not within the scope of this study; however, it should be emphasized that, as with journals, some conferences have stringent refereeing processes and low acceptance rates and others do not. All of the conference proceedings indexed by *Scopus* are peer-reviewed.

In conclusion, the findings suggest that many of the previous studies that used *WoS* exclusively to generate citation data to evaluate and/or rank scholars, journals, departments, and so on have been based on skewed and incomplete data and thus may have resulted in inaccurate assessments and imprecise rankings. One should, however, note that before 2005 *WoS* was really the only source available for conducting large-scale citation analyses. Future studies will have to employ both *WoS* and *Scopus* to generate accurate citation accounts of authors, journals, and academic programs because these two databases largely complement rather than replace each other.

Given the low overlap, or the high degree of uniqueness, in citations between the two databases, the findings further suggest that the use of *Scopus* in addition to *WoS* may have significant implications on the *h*-index scores of authors and journals (Bar-Ilan, 2006, submitted; Braun, Glänzel, & Schubert, 2006; Cronin & Meho, 2006; Kelly & Jennions, 2006; Oppenheim, 2007; Saad, 2006), journal impact factors (Garfield, 1996, 2006; Nisonger, 2004b), research assessment exercises as those conducted in the United Kingdom (Oppenheim, 1995; Smith & Eysenck, 2002; Warner, 2000), and on the correlation between citation data and perception-based evaluations or rankings (Meho & Sonnenwald, 2000; So, 1998). It will be important to verify the influence of *Scopus* data on these measures or practices as they are widely used by promotion and tenure committees, funding agencies, and collection development librarians, among others for assessing the research impact of scholars and departments and the quality of publications and journals.

**Effect of *Google Scholar* on Citation Counts and Rankings of LIS Faculty**

Data collected in this study show that, in contrast to *WoS* and *Scopus*, which index citations mainly from journal articles and conference papers, citations found through *GS* come from many different types of documents, including journal articles, conference papers, doctoral dissertations, master's theses, technical reports, research reports, chapters, and books, among others (see Table 8). Data also show that the majority of citations found through *GS* come from documents published after 1993 (see Table 9). A main reason for this is that the study group has less citable works published before 1993 in comparison to those published since then. Another reason is that, unlike *WoS* and *Scopus* which enter the citation information into their databases in a semi-automatic fashion, *GS* relies exclusively on the availability of online full text documents and, therefore, retrospective coverage will increase only as older materials are converted into digital format and published on the Web. As mentioned earlier, analysis in this study is based only on citations found in journal items and conference papers published between 1996 and 2005.

Results show that *GS* identifies 1,448 (or 53.0%) more citations than *WoS* and *Scopus* combined (4,181 citations for *GS* in comparison to 2,733 for the union of *WoS* and *Scopus*). Results also show that combining citations from *GS, WoS*, and *Scopus* increases the number of citations to SLIS faculty members as a whole by 93.4% (from 2,733 to 5,285 citations). In other words, one would miss over 93.4% of relevant citations if searching were limited to *WoS* and *Scopus*. While the high number of unique citations in GS could be very helpful for those individuals seeking promotion, tenure, faculty positions, or research grants, most of these citations come from low impact journals and/or conference proceedings (see below).

Data show that the percentage of increase in citation counts for SLIS faculty members varies considerably when *GS* results are added to those of *WoS* and *Scopus* (range=120.2%). Faculty members with research strengths in the areas of communities of practice, computer-mediated communication, data mining, data modeling, discourse analysis, gender and information technology, human-computer interaction, information retrieval, information visualization, knowledge discovery, and user modeling had their citation counts increase considerably more than those faculty members with research strengths in other areas (see Table 10). While this suggests that one should use *GS* to generate accurate citation counts of these authors, unlike the effect of adding *Scopus*' unique citations to those of *WoS*, adding *GS*'s unique citations data to those of *WoS* and *Scopus* does not significantly alter the relative ranking of faculty members—Spearman Rank Order correlation coefficient = 0.976 at 0.001 level (see Table 11).

Even when *GS* results are added to those of *WoS* and *Scopus* separately, *GS* results do not significantly change the relative ranking of scholars—the Spearman Rank Order correlation coefficient between *GS* and *WoS* = 0.874 and between *GS* and *Scopus* = .970. Perhaps equally important is that the overlap between *GS* and the union of *WoS* and *Scopus* is considerably low (30.8%) and that *GS* misses a high number (1,104 or 40.4%) of the 2,733 citations found by *WoS* and *Scopus* (see Figure 2). Both of these figures are very striking, especially given the fact that virtually all citations from *WoS* and *Scopus* come from refereed and/or reputable sources.

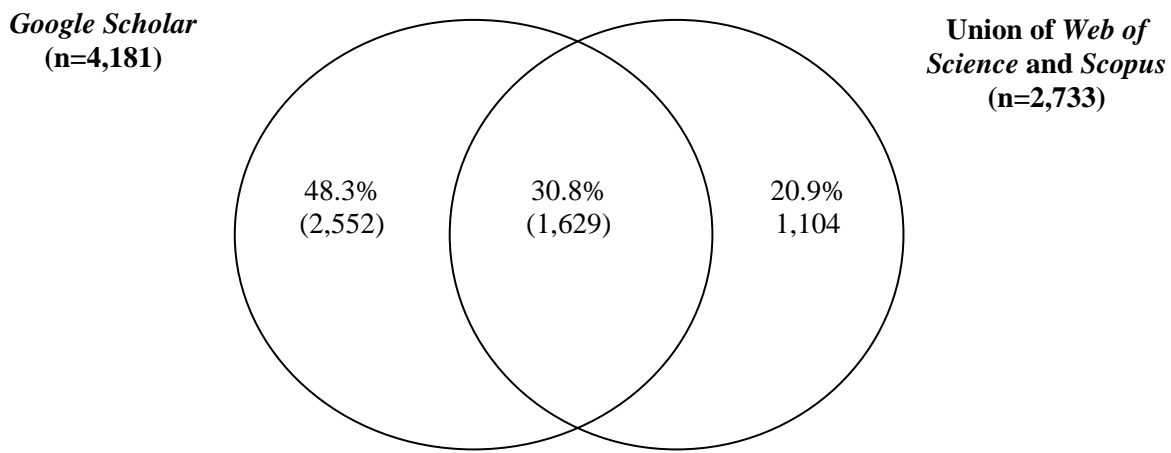

**Figure 2. Distribution of unique and overlapping citations in *GS* and *WoS* and *Scopus* (N=5,285)**

To test that these results were not an outcome of the study group size and source, citation data was collected for 10 additional LIS faculty members, specializing in several research areas such as archives, children and young adult librarianship, digital libraries, evaluation of library services, health and medical information, public libraries, and school library media. These 10 faculty members were identified through searches in *Library Literature and Information Science* and *WoS* databases and were selected based on the number of refereed journal articles each one of them had in these databases (at least 5 articles in one or both databases). Results of this group of faculty members, who are cited in 442 documents (333 journal articles, 68 conference papers, 29 review articles, 8 editorial materials, 3 book reviews, and 1 letter), showed very similar patterns to those of the main study group (see Table 12). The only major difference found was that *GS* increases the citation count of the test group by 51.9% in comparison to 93.4% for the main study group. This difference is mostly attributable to the fact that some of the main sources of citations covering the research areas of the test group are either not yet available in digital format (e.g., *The American Archivist* and *Archivaria*) or have only the last few years of their issues available online (e.g., *Archival Issues*, 1997-present; *Archival Science*, 2000-present; and *Journal of Education for Library and Information Science*, 2003-present). This provides evidence that *GS* favors those faculty members who work in research areas that have high web presence or those who make their scholarly works available online more than those faculty members who work in research areas that do not have high web presence or who do not post their works online.

Given the results from the test group as well as the fact that *GS* is so cumbersome to use, misses a significant number of citations from refereed sources, and has little or no influence on the relative rankings of scholars, one could conclude that, as far as LIS is concerned, *GS* is superfluous to using both *WoS* and *Scopus* to generate citation counts for assessing and comparing scholars, journals, and academic departments, especially when the focus of a study is on citations in refereed journals and conference proceedings. Results of this study also show that the use of Scopus in addition to WoS further diminishes the valued of using GS as evidenced in the increase in Spearman's *rho* correlation from .874 between *GS* and *WoS* to 0.976 between *GS* and the Union of *WoS* and *Scopus*.

Considering the type of documents in which the citations were found, *GS* retrieves significantly more (almost four times as many) citations from conference papers than *WoS* and *Scopus* combined (1,849 in comparison to 496, respectively). In contrast, *WoS* and *Scopus* retrieve almost as many citations from journals as *GS* does (1,968 in comparison to 2,332, respectively). It should be emphasized here that the relatively poor coverage of conference papers by *WoS* and *Scopus,* or the relative good coverage by *GS* of this document type, has much to do with the fact that many authors make their conference papers available online themselves. Almost half of *GS*'s unique citations from conference papers and many of its citations from journals were identified through full-text documents made available online by their authors (i.e., self-archived) rather than from the official web sites of the publishers of the conference proceedings and journals.[6] What this kind of findings reveals is that there is a dramatic advantage in favor of the articles that their authors make available online. According to Harnad and Brody (2004):

> The proportion of articles for which their authors provide OA [Open Access] is likely to increase dramatically now, in part because of the mounting evidence for *the impact advantage OA confers*. OA will also increase because of the growing number of journals that have already given their official "green light" to author self-archiving, partly because journal impact factors also benefit from increased article impact, and partly because journals are eager to demonstrate that they have no wish to stand in the way of OA and its benefits to research and researchers.

**Sources of Citations and Their Refereed Status and Language**

As mentioned earlier, only 58.2% (or 1,591) of all *WoS* and *Scopus* citations (n=2,733) were duplicated in both databases, raising the important question of where the 1,142 unique citations originated from. Answering this question will identify important LIS-relevant journals and conference proceedings that are not indexed by the databases. This list could be very useful for collection development librarians as well as for the producers of the databases should they decide to increase their coverage of LIS literature. Data show that the 2,023 citations from *WoS* come from 505 different journals and conference proceedings whereas the 2,301 citations from *Scopus* come from 681 different titles. The 2,733 citations from the union of both databases come from 787 different journals and conference proceedings. Of these 787 titles, 398 (or 50.5%) are indexed by both databases, 107 (or 13.5%) are indexed only by *WoS*, and 283 (or 36.0%) are indexed only by *Scopus*.

Data show that the top 54 (or 6.9%) sources of citations in *WoS* and/or *Scopus* account for 1,410 (or 51.6%) of all of the databases' 2,733 citations for the study group, reflecting the Matthew Effect in

---

[6]In these cases, data collection involved going to the root web site to identify the full bibliographic information of the citing documents. Most often, the root web sites were the curricula vitae of the authors of the citing documents.

citations—a small number of sources attracts the lion's share of citations and a large number of sources receives relatively few (Merton, 1968). Of these 54 sources of citations, 10 (or 18.5%) are not indexed by *WoS* whereas only one is not indexed by *Scopus*. Of the 10 titles not indexed by *WoS*, six are conference proceedings and four are journals (see Table 13). It is interesting to note that these four journals have higher impact factor scores than most of the LIS journals that are currently indexed by *WoS* and/or included in ISI's *Journal Citation Reports—Computer Supported Cooperative Work* (1.190), *Internet and Higher Education* (1.022), *D-Lib Magazine* (.615), and *International Review of Research in Open and Distance Learning* (.354), in comparison to a median impact factor of 0.480 for LIS journals (Thomson Scientific, 2006).[7] As for the six conference proceedings, three were from ACM (Association for Computing Machinery) and/or IEEE (Institute of Electrical and Electronics Engineers).

These results, which are influenced by the makeup of our study group (i.e., one with strong research focus in communities of practice, computer-mediated communication, human-computer interaction, and information visualization, in addition to traditional LIS research areas), suggest that if *WoS* is to reduce the gap in its coverage of LIS and LIS-related fields, it should consider adding at least the relevant high impact factor journals and conference proceedings that *Scopus* uniquely indexes. As is, the results imply that *Scopus* is necessary to use along with *WoS* for providing a better and more accurate picture of the impact LIS research makes on other fields, as evidenced by the computer science, education, and engineering titles that cite LIS literature and are only/primarily indexed by *Scopus*.

Further analysis shows that when a journal or a conference proceeding is indexed by both databases, *WoS* tends to identify more citations from these commonly indexed sources than *Scopus* does in the majority of cases. For example, *WoS* identifies 145 citations from the *Journal of the American Society for Information Science and Technology* whereas *Scopus* finds only 112 from the same journal covering the same time period (see Table 13 for more examples). There are, however, cases where *Scopus* identifies more citations than *WoS* from the same titles (e.g., *Journal of Computer-Mediated Communication—JCMC, Journal of Educational Computing Research,* and *Education for Information*). Reasons for these variations in coverage between the two databases include: database errors (e.g., lack of cited references information and incomplete lists of references in some database records), partial indexing of journal content (e.g., not indexing all articles published in a journal and not indexing book reviews as is the case in *Scopus* although some of these items contain citations), and incomplete coverage period of journals (e.g., missing an entire issue or volume—*JCMC*, for example is covered by *Scopus* from 1996 to present whereas it was just recently added to *WoS* covering the years 2005 to present).

---

[7]The 2006 Citation Impact Factor formula = number of citations in 2006 to articles published in 2005 + number of citations in 2006 to articles published in 2004, divided by number of articles published in 2004-2005.

As in the case of *Scopus*, *GS* results also raise the important question of where it found the 2,552 citations that were missed by both *WoS* and *Scopus*. As mentioned earlier, *GS* is able to search documents from hundreds of publishers, including items their authors themselves have made available online. An examination of the top 51 sources of citations found exclusively in *GS*, however, shows that 14 are actually indexed by *Scopus* and six are indexed by both *Scopus* and *WoS* (see Table 14). The reasons why *WoS* and *Scopus* miss some citations from these 20 titles are similar to those mentioned above (e.g., database errors, partial and incomplete coverage, etc.).

Results also show that 10 of the remaining top 51 sources of *GS* unique citations are journals and 21 are conference proceedings. To identify the quality or impact of these 31 titles, we used *Scopus* to generate citation counts to these titles and compared the counts to those of highly ranked LIS journals and conference proceedings, such as the *Annual Review of Information Science and Technology, College & Research Libraries, Information Processing & Management, Journal of Documentation, Journal of the American Society for Information Science and Technology,* and *Scientometrics* (Nisonger & Davis, 2005; Thomson Scientific, 2006). Results show that with the exception of one title, none of these 31 titles are cited more than the top 11 LIS journals and conference proceedings as identified by Nisonger and Davis (see Table 15). This finding raises important questions regarding the quality of citations uniquely found in *GS* as well as the wisdom of using these citations for tenure and promotion purposes, despite the fact that most of the citations uniquely found by *GS* are from refereed sources (only two of the top 51 sources are not refereed). Note that of the top 51 sources of citations, 15 are published or sponsored by ACM, three by IEEE, and three jointly by ACM and IEEE (see Table 14).

Another important finding is that *GS* provides significantly better coverage of non-English language materials (6.94% of its total citations) than both *WoS* (1.14%) and *Scopus* (0.70%) (see Table 16). This discovery suggests that *GS* is indispensable for showing one's international impact.

## CONCLUSIONS AND IMPLICATIONS

This study provides several useful suggestions for scholars conducting citation analysis and those who need assistance in compiling their own citation records. It informs researchers, administrators, editors, reviewers, funding agencies, and information professionals of novel ways of identifying citations to authors, papers, and journals. Until very recently, *WoS* was essentially the only practical source for locating citations. This study not only shows the value of, and need for, broadening the sources of citations, but also suggests that a significantly different map of scholarly networks could be developed when using multiple citation sources.

The study found that the addition of *Scopus* citations to those of *WoS* could significantly alter the ranking of scholars. The study also found that *GS* stands out in its coverage of conference proceedings as well as international, non-English language journals, among others. *GS* also indexes a wide variety of document types, some of which may be of significant value to researchers. The use of *Scopus* and *GS*, in addition to *WoS*, reveals a more comprehensive and accurate picture of the extent of the scholarly relationship between LIS and other fields, as evidenced by the unique titles that cite LIS literature (e.g., titles from Cognitive Science, Computer Science, Education, and Engineering, to name only a few). Significantly, this study has demonstrated that:

1) Although *WoS* remains an indispensable citation database, it should not be used alone for locating citations to an author or title, and, by extension, journals, departments, and countries; *Scopus* should be used concurrently.

2) Although *Scopus* provides more comprehensive citation coverage of LIS and LIS-related literature than *WoS* for the period 1996-2005, the two databases complement rather than replace each other.

3) While both *Scopus* and *GS* help identify a considerable number of citations not found in *WoS*, only *Scopus* significantly alters the ranking of scholars as measured by *WoS*.

4) Although *GS* unique citations are not of the same quality as those found in *WoS* or *Scopus*, they could be very useful in showing evidence of broader international impact than could possibly be done through the two proprietary databases.

5) *GS* value for citation searching purposes is severely diminished by its inherent problems. *GS* data are not limited to refereed, high quality journals and conference proceedings. *GS* is also very cumbersome to use and needs significant improvement in the way it displays search results and the downloading capabilities it offers for it to become a useful tool for large-scale citation analyses.

6) Given the low overlap or high uniqueness between the three tools, they may all be necessary to develop more accurate maps or visualizations of scholarly networks and impact both within and between disciplines (Börner, Chen, & Boyack, 2003; Börner, Sanyal, & Vespignani, 2006; Small, 1999; White & McCain, 1997).

7) Each database or tool requires specific search strategy(ies) in order to collect citation data, some more accurately and quickly (i.e., *WoS* and *Scopus*) than others (i.e., *GS*).

This study has significant implications for funding agencies as well as editors and publishers of journals who may wish to use citation counts and rankings to identify subject experts to review grant applications or submitted manuscripts and to determine the impact of projects and articles they funded or published. The study has also significant implications for the wider scholarly community as researchers begin to adopt the methods and databases described or listed here to identify citations that may otherwise remain unknown. Continuous advances in information technology and improvements in online access to citation data suggest that future studies should explore:

- Databases and tools that can be used to locate citations from refereed sources not covered by *WoS* or *Scopus*;
- Potential impact of these databases and tools on citation counts as well as the correlation between citation counts and peer reviews/assessments of publication venues;
- Whether broader sourcing of citations alters a paper, author, journal, or a department's relative ranking vis-à-vis others and, if so, how;
- Which sources of citations provide better coverage of certain research areas than others; and
- The intrinsic quality of citations found in these sources.

The recent emergence of *Scopus* and other citation databases and tools has certainly marked the beginning of a new era in citation and bibliometrics analyses. If *WoS* wants to improve its quality and fend off competition from such rivals as *Scopus* and perhaps *GS*, it needs to broaden its coverage by indexing more high impact journals and conference proceedings.

**ACKNOWLEDGMENTS**

We are most grateful to Shahriar Akram, Kathleen Burlingame, Sara Franks, Lana Gottschalk, Sarah Horowitz, Mary Snyder, and Jacqueline Steinhofer for assistance with data collection and to Blaise Cronin, Peter Jacsó, Alice Robbin, Debora Shaw, and an anonymous reviewer for helpful comments on the paper.

**NOTE**

This project was partially funded by OCLC Library and Information Science Research Grant Project.

**CITED REFERENCES**

**Table 1. Comparisons of Databases and Tools Used in the Study**

|  | **Web of Science** | **Scopus** | **Google Scholar** |
|---|---|---|---|
| **Breadth of coverage** | 36 million records (1955-) 8,700 titles (including 190 open access journals and several hundred conference proceedings) | 28 million records (1966-) 15,000 titles (including 12,850 journals, 700 conference proceedings, 600 trade publications, 500 open access journals, and 125 book series) | Unknown number of records Unknown number of sources Over 30 different document types Unknown number of publishers |
| **Depth of coverage** | A&HCI: 1975- SCI: 1900- SSCI: 1956- | With cited references data: 1996- Without cited references data: 1966- | Unknown |
| **Subject coverage** | All | All | All |
| **Citation browsing options** | Cited author Cited work | Not available | Not available |
| **Citation searching options** | Cited author Cited work (requires use of the abbreviated journal, book, or conference title in which the work appeared) Cited year | The "Basic Search" interface allows keyword and phrase searching via "References" field. The "Advanced Search" interface, allows searching for: Cited author (REFAUTH) Cited title (REFTITLE) Cited work (REFSRCTITLE) Cited year (REFPUBYEAR) Cited page (REFPAGE) Cited reference (REF), which is a combined field that searches the REFAUTH, REFTITLE, REFSRCTITLE, REFPUBYEAR, and REFPAGE fields. | Keyword and phrase searching Limit/search options include: "Author," "Publication," "Date," and "Subject Areas" |
| **Analytical tools** | Ranking by author, publication year, source name, country, institution name, subject category, language, and document type | Ranking by author, publication year, source name, subject category, and document type. Analysis of citations by year (via Citation Tracker) | Not available |
| **Downloading and exporting options to citation management software (e.g., EndNote and RefWorks)** | Yes | Yes | Yes |

**Table 2. SLIS Publication Data (1970-2005)***

| Document Type | Count** |
|---|---|
| Refereed journal articles | 322 (312) |
| Conference papers | 313 (305) |
| Chapters | 135 (131) |
| Non-refereed journal articles | 94 (93) |
| Technical reports / Working papers | 84 (83) |
| Articles in professional journals | 59 (59) |
| Books | 36 (36) |
| Edited volumes | 35 (35) |
| Bibliographies | 16 (16) |
| Review articles | 15 (12) |
| Encyclopedia articles | 11 (11) |
| **Total** | **1,120 (1,093)** |

*Book reviews, meeting abstracts, letters to editors, panels, presentations (invited and otherwise), and so on are excluded from this table.
**Figures in parentheses refer to unique records (i.e., after removing duplicates due to co-authorship among SLIS faculty members).

**Table 3. Citation Count by Year – *Web of Science* and *Scopus***

| Years | *WoS* | *Scopus* | Union of *WoS* and *Scopus* |
|---|---|---|---|
| 1971-1975 | 1 | - | 1 |
| 1976-1980 | 15 | - | 15 |
| 1981-1985 | 129 | - | 129 |
| 1986-1990 | 201 | - | 201 |
| 1991-1995 | 323 | - | 323 |
| **Subtotal** | **669** | **-** | **669** |
| 1996 | 119 | 101 | 140 |
| 1997 | 121 | 119 | 144 |
| 1998 | 142 | 123 | 167 |
| 1999 | 131 | 128 | 164 |
| 2000 | 175 | 171 | 219 |
| 2001 | 207 | 242 | 278 |
| 2002 | 202 | 220 | 271 |
| 2003 | 251 | 291 | 348 |
| 2004 | 323 | 459 | 510 |
| 2005 | 352 | 447 | 492 |
| **Subtotal** | **2,023** | **2,301** | **2,733** |
| **TOTAL** | **2,692** | **2,301** | **3,402** |

**Table 4. Impact of Adding *Scopus* Citations on Faculty and School Citation Counts (1996-2005)**

| Research areas of individual faculty members* | WoS | Scopus | Union of WoS and Scopus | % Increase |
|---|---|---|---|---|
| Human-computer interaction | 544 | 740 | 853 | 56.80% |
| Citation analysis, informetrics, scholarly communication, and strategic intelligence | 508 | 459 | 564 | 11.00% |
| Computer-mediated communication, gender and information technology, and discourse analysis | 273 | 313 | 365 | 33.70% |
| E-commerce, information architecture, information policy and electronic networking | 162 | 168 | 188 | 16.00% |
| Bibliometrics, Collection development and management, evaluation of library sources and services, and serials | 123 | 108 | 137 | 11.40% |
| Information seeking and use, design and impact of electronic information sources, and informetrics | 122 | 111 | 128 | 4.90% |
| Intelligent interfaces for information retrieval and filtering, knowledge discovery, and user modeling | 118 | 129 | 154 | 30.50% |
| Information visualization, data mining, and data modeling | 115 | 133 | 165 | 43.50% |
| Communities of practice | 88 | 159 | 175 | 98.90% |
| Classification and categorization, ontologies, metadata, and information architecture | 83 | 80 | 93 | 12.00% |
| Critical theory and documentation | 35 | 37 | 42 | 20.00% |
| Computational linguistics, computer-mediated communication, and sociolinguistics and language acquisition | 32 | 38 | 44 | 37.50% |
| Citation analysis, bibliometrics, and data retrieval and integration | 29 | 21 | 31 | 6.90% |
| Information retrieval and data integration | 28 | 32 | 40 | 42.90% |
| Information policy, social and organizational informatics, and research methods | 28 | 31 | 34 | 21.40% |
| **Faculty Members Total** | **2,288** | **2,559** | **3,013** | 31.70% |
| **School Total**** | **2,023** | **2,301** | **2,733** | 35.10% |

*Each row in the table represents a single faculty member and the main research topics covered by him/her. It would have been practically impossible to classify citations by individual topics rather than individual faculty members.
**Excludes duplicate citations.

**Table 5. Impact of Adding *Scopus* Citations on the Ranking of Faculty Members (1996-2005)**

| Faculty Member* | *WoS* | | Union of *WoS* and *Scopus* | |
|---|---|---|---|---|
| | Count | Rank | Count | Rank |
| A | 544 | 1 | 853 | 1 |
| B | 508 | 2 | 564 | 2 |
| C | 273 | 3 | 365 | 3 |
| D | 162 | 4 | 188 | 4 |
| E | 123 | 5 | 137 | 8 |
| F | 122 | 6 | 128 | 9 |
| G | 118 | 7 | 154 | 7 |
| H | 115 | 8 | 165 | 6 |
| I | 88 | 9 | 175 | 5 |
| J | 83 | 10 | 93 | 10 |
| K | 35 | 11 | 42 | 12 |
| L | 32 | 12 | 44 | 11 |
| M | 29 | 13 | 31 | 15 |
| N | 28 | 14T | 40 | 13 |
| O | 28 | 14T | 34 | 14 |

**Table 6. Overlap Between *Scopus* and *Web of Science* (1996-2005)**

| Research areas of individual faculty members* | *WoS* | *Scopus* | Union | Overlap | % |
|---|---|---|---|---|---|
| Human-computer interaction | 544 | 740 | 853 | 430 | 50.4% |
| Citation analysis, informetrics, scholarly communication, and strategic intelligence | 508 | 459 | 564 | 403 | 71.5% |
| Computer-mediated communication, gender and information technology, and discourse analysis | 273 | 313 | 365 | 221 | 60.5% |
| E-commerce, information architecture, information policy and electronic networking | 162 | 168 | 188 | 142 | 75.5% |
| Bibliometrics, Collection development and management, evaluation of library sources and services, and serials | 123 | 108 | 137 | 94 | 68.6% |
| Information seeking and use, design and impact of electronic information sources, and informetrics | 122 | 111 | 128 | 105 | 82.0% |
| Intelligent interfaces for information retrieval and filtering, knowledge discovery, and user modeling | 118 | 129 | 154 | 83 | 53.9% |
| Information visualization, data mining, and data modeling | 115 | 133 | 165 | 92 | 55.8% |
| Communities of practice | 88 | 159 | 175 | 72 | 41.1% |
| Classification and categorization, ontologies, metadata, and information architecture | 83 | 80 | 93 | 70 | 75.3% |
| Critical theory and documentation | 35 | 37 | 42 | 30 | 71.4% |
| Computational linguistics, computer-mediated communication, and sociolinguistics and language acquisition | 32 | 38 | 44 | 26 | 59.1% |
| Citation analysis, bibliometrics, and data retrieval and integration | 29 | 21 | 31 | 19 | 61.3% |
| Information retrieval and data integration | 28 | 32 | 40 | 20 | 50.0% |
| Information policy, social and organizational informatics, and research methods | 28 | 31 | 34 | 25 | 73.5% |
| **Faculty Members Total** | **2,288** | **2,559** | **3,013** | **1,832** | 60.8% |
| **School Total**** | **2,023** | **2,301** | **2,733** | **1,591** | 58.2% |

*Each row in the table represents a single faculty member and the main research topics covered by him/her. It would have been practically impossible to classify citations by individual topics rather than individual faculty members.
**Excludes duplicate citations.

**Table 7.** *Web of Science* and *Scopus* Citation Count by Document Type (1996-2005)

| Document Type | WoS | | Scopus | | Union | |
|---|---|---|---|---|---|---|
| | Count* | % | Count* | % | Count* | % |
| Journal articles | 1,529 | 75.6% | 1,754 | 76.2% | 1,968 | 72.0% |
| Conference papers | 229 | 11.3% | 359 | 15.6% | 496 | 18.1% |
| Review articles | 172 | 8.5% | 147 | 6.4% | 175 | 6.4% |
| Editorial materials | 63 | 3.1% | 36 | 1.6% | 64 | 2.3% |
| Book reviews | 17 | 0.8% | 0 | 0.0% | 17 | 0.6% |
| Letters to editors | 9 | 0.4% | 2 | 0.1% | 9 | 0.3% |
| Bibliographic essays | 2 | 0.1% | 2 | 0.1% | 2 | 0.1% |
| Biographical item | 2 | 0.1% | 1 | 0.0% | 2 | 0.1% |
| **Total** | **2,023** | **100.0%** | **2,301** | **100.0%** | **2,733** | **100.0%** |
| Total from Journals | 1,794 | 88.7% | 1,942 | 84.4% | 2,237 | 81.9% |
| Total from Conference Papers | 229 | 11.3% | 359 | 15.6% | 496 | 18.1% |
| **Total** | **2,023** | **100.0%** | **2,301** | **100.0%** | **2,733** | **100.0%** |

*Excludes duplicate citations.

**Table 8.** *Google Scholar* Citation Count by Document Type (1996-2005)*

| Document Type | Count** | % |
|---|---|---|
| Journal Articles | 2,215 | 40.32% |
| Conference Papers | 1,849 | 33.66% |
| Doctoral Dissertations | 261 | 4.75% |
| Master's Theses | 243 | 4.42% |
| Book Chapters | 199 | 3.62% |
| Technical Reports | 129 | 2.35% |
| Reports | 110 | 2.00% |
| Books | 102 | 1.86% |
| Review Articles | 86 | 1.57% |
| Conference Presentations | 72 | 1.31% |
| Unpublished Papers | 65 | 1.18% |
| Bachelor's Theses | 34 | 0.62% |
| Working Papers | 31 | 0.56% |
| Editorial Materials | 25 | 0.46% |
| Research Reports | 23 | 0.42% |
| Workshop Papers | 15 | 0.27% |
| Doctoral Dissertation Proposals | 9 | 0.16% |
| Conference Posters | 9 | 0.16% |
| Book Reviews | 3 | 0.05% |
| Master's Thesis Proposals | 3 | 0.05% |
| Preprints | 3 | 0.05% |
| Conference Paper Proposals | 2 | 0.04% |
| Government Documents | 2 | 0.04% |
| Letters to the editor | 2 | 0.04% |
| Biographical Item | 1 | 0.02% |
| **Total** | **5,493** | **100.00%** |
| Total from Journals | 2,332 | 42.45% |
| Total from Conference Papers | 1,849 | 33.66% |
| **Total from Journals and Conference Papers** | **4,181** | **76.12%** |
| Total from Dissertations/Theses | 538 | 9.79% |
| Total from Books | 301 | 5.48% |
| Total from Reports | 262 | 4.77% |
| Total from other document types | 211 | 3.84% |
| **Total** | **5,493** | **100.00%** |

*Table excludes 475 citations that did not have complete bibliographic information. These citations included: bachelor theses, presentations, grant proposals, doctoral qualifying examinations, submitted manuscripts, syllabi, term papers, research proposals, working papers, web documents, preprints, student portfolios, and so on.
**Excludes duplicate citations.

**Table 9.** *Google Scholar* **Citation Distribution by Year**

| Years | Citations from Journals and Conference Papers |
|---|---|
| 1971-1975 | 1 |
| 1976-1980 | 1 |
| 1981-1985 | 9 |
| 1986-1990 | 29 |
| 1991 | 4 |
| 1992 | 12 |
| 1993 | 9 |
| 1994 | 43 |
| 1995 | 67 |
| **Subtotal** | **175** |
| 1996 | 101 |
| 1997 | 145 |
| 1998 | 176 |
| 1999 | 248 |
| 2000 | 350 |
| 2001 | 409 |
| 2002 | 539 |
| 2003 | 671 |
| 2004 | 752 |
| 2005 | 790 |
| **Subtotal** | **4,181** |
| **TOTAL** | **4,356** |

**Table 10. Impact of Adding *Google Scholar* Citations on Faculty Members' Citation Count (1996-2005)**

| Research areas of individual faculty members* | Union of *WoS* and *Scopus* | GS | Union of Three Sources | % Increase |
|---|---|---|---|---|
| Human-computer interaction | 853 | 1,786 | 2,078 | **143.6%** |
| Citation analysis, informetrics, scholarly communication, and strategic intelligence | 564 | 517 | 802 | **42.2%** |
| Computer-mediated communication, gender and information technology, and discourse analysis | 365 | 671 | 797 | **118.4%** |
| E-commerce, information architecture, information policy and electronic networking | 188 | 164 | 244 | **29.8%** |
| Bibliometrics, Collection development and management, evaluation of library sources and services, and serials | 137 | 94 | 169 | **23.4%** |
| Information seeking and use, design and impact of electronic information sources, and informetrics | 128 | 114 | 171 | **33.6%** |
| Intelligent interfaces for information retrieval and filtering, knowledge discovery, and user modeling | 154 | 260 | 291 | **89.0%** |
| Information visualization, data mining, and data modeling | 165 | 187 | 249 | **50.9%** |
| Communities of practice | 175 | 342 | 403 | **130.3%** |
| Classification and categorization, ontologies, metadata, and information architecture | 93 | 76 | 125 | **34.4%** |
| Critical theory and documentation | 42 | 46 | 60 | **42.9%** |
| Computational linguistics, computer-mediated communication, and sociolinguistics and language acquisition | 44 | 73 | 92 | **109.1%** |
| Citation analysis, bibliometrics, and data retrieval and integration | 31 | 29 | 39 | **25.8%** |
| Information retrieval and data integration | 40 | 46 | 59 | **47.5%** |
| Information policy, social and organizational informatics, and research methods | 34 | 20 | 42 | **23.5%** |
| **Faculty Members Total** | **3,013** | **4,425** | **5,621** | **86.6%** |
| **School Total**** | **2,733** | **4,181** | **5,285** | **93.4%** |

Included in this table are citations from journals and conference papers only. Excluded are citations from dissertations, theses, reports, books, conference presentations, meeting abstracts, research and technical reports, unpublished papers, working papers, workshop papers, and so on. Table also excludes 475 citations that did not have complete bibliographic information. These citations included: bachelor theses, presentations, grant proposals, doctoral qualifying examinations, submitted manuscripts, syllabi, term papers, research proposals, working papers, web documents, preprints, student portfolios, and so on.

*Each row in the table represents a single faculty member and the main research topics covered by him/her. It would have been practically impossible to classify citations by individual topics rather than individual faculty members.

**Excludes duplicate citations.

**Table 11. Impact of Adding *Google Scholar* Citations on the Ranking of Faculty Members (1996-2005)**

| Faculty Member | Union of *WoS* and *Scopus* | | Union of *WoS*, *Scopus*, and *GS* | |
|---|---|---|---|---|
| | Count | Rank | Count | Rank |
| A | 853 | 1 | 2,078 | 1 |
| B | 564 | 2 | 802 | 2 |
| C | 365 | 3 | 797 | 3 |
| D | 188 | 4 | 244 | 7 |
| I | 175 | 5 | 402 | 4 |
| H | 165 | 6 | 249 | 6 |
| G | 154 | 7 | 291 | 5 |
| E | 137 | 8 | 169 | 9 |
| F | 128 | 9 | 171 | 8 |
| J | 93 | 10 | 125 | 10 |
| L | 44 | 11 | 92 | 11 |
| K | 42 | 12 | 60 | 12 |
| N | 40 | 13 | 59 | 13 |
| O | 34 | 14 | 42 | 14 |
| M | 31 | 15 | 39 | 15 |

**Table 12. Comparison of Results between Main Study Group and Test Group (1996-2005)**

| | Main Study Group | Test Group |
|---|---|---|
| Ratio of citations between *WoS* and *Scopus* | .88 : 1.0 | .91 : 1.0 |
| Percent of increase when adding *Scopus* unique citations to those of *WoS* | 35.1% | 40.6% |
| Percent of overlap between *WoS* and *Scopus* | 58.2% | 50.5% |
| Percent of unique citations found in *Scopus* in the union of *WoS* and *Scopus* | 26.0% | 27.8% |
| Percent of unique citations found in *WoS* in the union of *WoS* and *Scopus* | 15.8% | 21.6% |
| Percent of increase when adding *GS* unique citations to those of *WoS* and *Scopus* | 93.4% | 51.9% |
| Percent of overlap between *GS* and union of *WoS* and *Scopus* | 30.8% | 26.2% |
| Percent of *WoS* and *Scopus* citations missed by *GS* | 40.4% | 60.8% |
| Spearman Rank Order correlation coefficient between *GS* and union of *WoS* and *Scopus* | .976 | .967 |

Spearman Rank Order correlation coefficient between *GS* and union of *WoS* and *Scopus* for all 25 faculty members (i.e., main study group and test group combined) = .982.

**Table 13. Sources of Citations in *Scopus* and *Web of Science* (1996-2005)***

| Title | WoS | Scopus | Union | Rank | % (n=2,733) |
|---|---|---|---|---|---|
| Journal of the American Society for Information Science and Technology | 145 | 112 | 147 | 1 | 5.4% |
| Lecture Notes in Computer Science series** | 118 | 33 | 118 | 2 | 4.3% |
| Scientometrics | 78 | 69 | 79 | 3 | 2.9% |
| Journal of Documentation | 71 | 56 | 71 | 4 | 2.6% |
| Annual Review of Information Science and Technology | 53 | 51 | 53 | 5 | 1.9% |
| Journal of Information Science | 46 | 47 | 47 | 6 | 1.7% |
| Proceedings of the Annual Meeting of the American Society for Information Science and Technology | 41 | 35 | 45 | 7 | 1.6% |
| Journal of Computer-Mediated Communication | 15 | 41 | 41 | 8 | 1.5% |
| International Journal of Human-Computer Studies | 40 | 40 | 40 | 9 | 1.5% |
| Lecture Notes in Artificial Intelligence series** | 40 | 7 | 40 | 10 | 1.5% |
| Information Processing & Management | 39 | 38 | 39 | 11 | 1.4% |
| Interacting With Computers | 34 | 34 | 34 | 12 | 1.2% |
| Library & Information Science Research | 31 | 27 | 32 | 13 | 1.2% |
| Information Society: An International Journal | 24 | 25 | 28 | 14 | 1.0% |
| Aslib Proceedings | 22 | 14 | 23 | 15 | 0.8% |
| Behaviour and Information Technology | 22 | 23 | 23 | 16 | 0.8% |
| Computers & Education | 22 | 21 | 22 | 17 | 0.8% |
| College & Research Libraries | 21 | 17 | 21 | 18 | 0.8% |
| Library Trends | 21 | 21 | 21 | 19 | 0.8% |
| Computers in Human Behavior | 20 | 20 | 20 | 20 | 0.7% |
| Information Research: An International Electronic Journal | 16 | 19 | 20 | 21 | 0.7% |
| Journal of Academic Librarianship | 19 | 18 | 20 | 22 | 0.7% |
| Cyberpsychology & Behavior | 18 | 18 | 19 | 23 | 0.7% |
| Internet Research | 19 | 16 | 19 | 24 | 0.7% |
| Computer Supported Cooperative Work: The Journal of Collaborative Computing | Not indexed | 18 | 18 | 25 | 0.7% |
| Internet and Higher Education | Not indexed | 18 | 18 | 26 | 0.7% |
| Knowledge Organization | 18 | 15 | 18 | 27 | 0.7% |
| ACM Conference on Computer Supported Cooperative Work (CSCW) | Not indexed | 16 | 16 | 28 | 0.6% |
| Journal of Educational Computing Research | 4 | 16 | 16 | 29 | 0.6% |
| Library Collections, Acquisitions, and Technical Services | 16 | 10 | 16 | 30 | 0.6% |
| Human-Computer Interaction | 15 | 14 | 15 | 31 | 0.5% |
| SIGCHI Conference on Human Factors in Computing Systems (ACM) | Not indexed | 15 | 15 | 32 | 0.5% |
| Journal of Computer Assisted Learning | 14 | 14 | 14 | 33 | 0.5% |
| Libri | 14 | Not indexed | 14 | 34 | 0.5% |
| SPIE Proceedings Series (International Society for Optical Engineering) | Not indexed | 14 | 14 | 35 | 0.5% |
| Education for Information | 2 | 11 | 13 | 36 | 0.5% |
| ETR&D-Educational Technology Research & Development | 12 | 11 | 13 | 37 | 0.5% |
| Library Quarterly | 13 | 12 | 13 | 38 | 0.5% |
| Proceedings of the National Academy of Sciences | 13 | 13 | 13 | 39 | 0.5% |
| Government Information Quarterly | 11 | 10 | 12 | 40 | 0.4% |
| Library Resources & Technical Services | 12 | 10 | 12 | 41 | 0.4% |
| Online Information Review | 12 | 11 | 12 | 42 | 0.4% |
| American Society for Engineering Education (ASEE) Annual Conference | Not indexed | 11 | 11 | 43 | 0.4% |
| Canadian Journal of Information and Library Science | 10 | 11 | 11 | 44 | 0.4% |
| D-Lib Magazine | Not indexed | 11 | 11 | 45 | 0.4% |
| International Review of Research in Open and Distance Learning | Not indexed | 11 | 11 | 46 | 0.4% |
| Language Learning & Technology | 9 | 11 | 11 | 47 | 0.4% |
| Serials Librarian | 9 | 10 | 11 | 48 | 0.4% |
| ACM/IEEE Joint Conference on Digital Libraries / ACM Conference on Digital Libraries | Not indexed | 10 | 10 | 49 | 0.4% |
| Annual Meeting of the Human Factors and Ergonomics Society | Not indexed | 10 | 10 | 50 | 0.4% |
| Bioinformatics | 9 | 10 | 10 | 51 | 0.4% |
| Educational Technology & Society | 5 | 10 | 10 | 52 | 0.4% |
| International Journal of Information Management | 10 | 8 | 10 | 53 | 0.4% |
| Journal of the American Medical Informatics Association | 10 | 9 | 10 | 54 | 0.4% |
| **Total** | **1,193** | **1,152** | **1,410** | | **51.6%** |

*Highly cited titles discovered through the test group but not through the main study group include: *School Library Media Research* (14), *Medical Reference Services Quarterly* (9), *Archival Science* (8), and *OCLC Systems & Services* (6). All four are indexed by, or found through, *Scopus* and *GS*, but not *WoS*.

**Citations published in *LNCS* and *LNAI* series come from a number of different conference proceedings, such as: Advances in Case-Based Reasoning: European Workshop, CHI conferences, International Conference on Case-Based Reasoning Research and Development, European Conference on Research and Advanced Technology for Digital Libraries, and International Conference on Web-Based Learning.

Title changes of journals and conference proceedings were taken into consideration in the calculation of citation count.

**Table 14. Sources of Citations UNIQUE to *Google Scholar* (1996-2005)**

| SOURCE | Status | Count |
|---|---|---|
| **Annual Hawaii International Conference on System Sciences (HICSS) (IEEE)** | **Refereed** | **57** |
| Ciencia da Informacao | Refereed | 35 |
| CHI Extended Abstracts on Human Factors in Computing Systems (ACM) | Refereed | 30 |
| **SIGCHI Conference on Human Factors in Computing Systems (ACM)** | **Refereed** | **30** |
| HCI International: International Conference on Human-Computer Interaction | Refereed | 27 |
| Conference on Interaction Design and Children (ACM) | Refereed | 23 |
| OZCHI: Australian Conference on Computer-Human Interaction (IEEE) | Refereed | 23 |
| **Lecture Notes in Computer Science (ACM, IEEE, and others)** | **Refereed** | **22** |
| Annual Conference of the Australasian Society for Computers in Learning in Tertiary Education – ASCILITE | Refereed | 21 |
| Interact: IFIP TC13 International Conference on Human Computer Interaction | Refereed | 20 |
| European Conference on Computer Supported Cooperative Work | Refereed | 19 |
| Information Systems Research Seminar in Scandinavia (IRIS) | Refereed | 18 |
| **ACM Conference on Computer Supported Cooperative Work** | **Refereed** | **17** |
| Australasian Journal of Educational Technology / Australian Journal of Educational Technology | Refereed | 15 |
| **International ACM SIGGROUP Conference on Supporting Group Work (Group)** | **Refereed** | **14** |
| Internet Research Annual: Association of Internet Researchers (AOIR) Annual Conference | Refereed | 14 |
| Nordic Conference on Human-Computer Interaction | Refereed | 14 |
| **Personal and Ubiquitous Computing (ACM)** | **Refereed** | **14** |
| International Conference on Human Computer Interaction with Mobile Devices and Services (Mobile HCI) (ACM) | Refereed | 13 |
| **Proceedings of the Annual Meeting of the American Society for Information Science and Technology** | **Refereed** | **13** |
| **ACM/IEEE Joint Conference on Digital Libraries / ACM Conference on Digital Libraries** | **Refereed** | **12** |
| Computer Assisted Language Learning | Refereed | 11 |
| Americas Conference on Information Systems (AIS) | Refereed | 10 |
| Education and Information Technologies | Refereed | 10 |
| IFLA General Conference | Not refereed | 10 |
| Text Retrieval Conference (TREC) | Not refereed | 10 |
| ACM SIGGROUP Bulletin | Refereed | 9 |
| **IEEE International Conference on Information Visualisation** | **Refereed** | **9** |
| Interactions (ACM) | Refereed | 9 |
| International Conference on Artificial Intelligence in Education (AIED) | Refereed | 9 |
| **Journal of the American Society for Information Science and Technology** | **Refereed** | **9** |
| Annual Meeting of the American Educational Research Association (AERA) | Refereed | 8 |
| **Cognition** | **Refereed** | **8** |
| Conference on Designing Interactive Systems: Processes, Practices, Methods, and Techniques (DIS) (ACM) | Refereed | 8 |
| **Distance Education: An International Journal** | **Refereed** | **8** |
| **International Conference on Ubiquitous Computing (UbiComp)** | **Refereed** | **8** |
| ACM International Conference of Computer-Supported Collaborative Learning (CSCL) | Refereed | 7 |
| **ACM/IEEE International Conference on Software Engineering (ICSE)** | **Refereed** | **7** |
| Educational Media International | Refereed | 7 |
| Ethics and Information Technology | Refereed | 7 |
| ACM SIGCHI Bulletin | Refereed | 6 |
| **ACM Transactions on Computer-Human Interaction** | **Refereed** | **6** |
| American Journal of Distance Education | Refereed | 6 |
| **Annual International ACM SIGIR Conference on Research and Development in Information Retrieval** | **Refereed** | **6** |
| **Collection Building** | **Refereed** | **6** |
| Ed-Media: World Conference on Educational Multimedia | Refereed | 6 |
| **Interacción** | **Refereed** | **6** |
| **Issues in Information Systems** | **Refereed** | **6** |
| **Journal of Educational Technology Systems** | **Refereed** | **6** |
| Latin American Conference on Human-Computer Interaction | Refereed | 6 |
| Perspectivas em Ciência da Informação | Refereed | 6 |

Titles in bold are journals/conference proceedings indexed by *WoS* and/or *Scopus*. Title changes of journals and conference proceedings were taken into consideration in the calculation of citation count.

**Table 15. Comparison Between Citation Counts of Top 46 Sources of *Google Scholar* Unique Citations and Those of Top LIS Journals and Conference Proceedings**

| TITLE | Citation Count | Rank |
|---|---|---|
| Journal of the American Society for Information Science and Technology | 6,679 | 1 |
| Information Processing and Management | 5,164 | 2 |
| **Annual Meeting of the American Educational Research Association (AERA)** | **4,357** | **3** |
| Journal of Information Science | 2,917 | 4 |
| Scientometrics | 2,596 | 5 |
| Journal of Documentation | 2,507 | 6 |
| Proceedings of the Annual Meeting of the American Society for Information Science and Technology | 1,654 | 7 |
| Journal of Academic Librarianship | 1,459 | 8 |
| Annual Review of Information Science and Technology | 1,349 | 9 |
| Library Quarterly | 1,268 | 10 |
| College & Research Libraries | 1,197 | 11 |
| Library & Information Science Research | 1,023 | 12 |
| **European Conference on Computer Supported Cooperative Work** | **872** | **13** |
| **Ed-Media: World Conference on Educational Multimedia** | **866** | **14** |
| **International Conference on Ubiquitous Computing (UbiComp)** | **824** | **15** |
| **HCI International: International Conference on Human-Computer Interaction** | **783** | **16** |
| **International Conference of Computer-Supported Collaborative Learning (CSCL) (ACM)** | **703** | **17** |
| **American Journal of Distance Education** | **684** | **18** |
| **Americas Conference on Information Systems (AIS)** | **637** | **19** |

Source of citation count data for this table: *Scopus* (1996-2006).
All of the non-bold items are either fully or selectively indexed by *WoS* and/or *Scopus* and are top ranked LIS journals and conference proceedings, according to *Journal Citation Reports* (Thomson Scientific, 2006) and Nisonger and Davis (2005). All of the bold items are sources uniquely searched/identified by *GS*.
Title changes of journals and conference proceedings were taken into consideration in the calculation of citation count.

**Table 16. Citation Count Distribution by Language (1996-2005)**

|  | *WoS* | | *Scopus* | | GS | | Total | |
|---|---|---|---|---|---|---|---|---|
|  | **Count** | **%** | **Count** | **%** | **Count*** | **%** | **Count*** | **%** |
| English | 2,000 | 98.86% | 2,285 | 99.30% | 3,891 | 93.06% | 4,972 | 94.08% |
| Portuguese |  |  |  |  | 92 | 2.20% | 92 | 1.74% |
| Spanish | 4 | 0.20% | 3 | 0.13% | 63 | 1.51% | 68 | 1.29% |
| German | 13 | 0.64% | 9 | 0.39% | 38 | 0.91% | 50 | 0.95% |
| Chinese |  |  |  |  | 44 | 1.05% | 44 | 0.83% |
| French | 3 | 0.15% | 1 | 0.04% | 32 | 0.77% | 35 | 0.66% |
| Italian |  |  |  |  | 8 | 0.19% | 8 | 0.15% |
| Japanese | 3 | 0.15% | 3 | 0.13% | 1 | 0.02% | 4 | 0.08% |
| Swedish |  |  |  |  | 3 | 0.07% | 3 | 0.06% |
| Czech |  |  |  |  | 2 | 0.05% | 2 | 0.04% |
| Dutch |  |  |  |  | 2 | 0.05% | 2 | 0.04% |
| Finnish |  |  |  |  | 2 | 0.05% | 2 | 0.04% |
| Croatian |  |  |  |  | 1 | 0.02% | 1 | 0.02% |
| Hungarian |  |  |  |  | 1 | 0.02% | 1 | 0.02% |
| Polish |  |  |  |  | 1 | 0.02% | 1 | 0.02% |
| **Total** | **2,023** | **100.00%** | **2,301** | **100.00%** | **4,181** | **100.00%** | **5,285** | **100.00%** |
| **Non-English** | **23** | **1.14%** | **16** | **0.70%** | **290** | **6.94%** | **313** | **5.92%** |